# Phase Change Induced Magnetic Switching through Metal-insulator Transition in VO$_2$/TbFeCo Films

Chung T. Ma [1], Salinporn Kittiwatnakul[1,2], Apiprach Sittipongpittaya [1], Yuhan Wang [1], Md Golam Morshed [3], Avik W. Ghosh [1,3], and S. Joseph Poon [*,4,1]

[1] Department of Physics, University of Virginia, Charlottesville, Virginia, 22904 USA
[2] Department of Physics, Faculty of Science, Chulalongkorn University, 10330 Bangkok Thailand
[3] Department of Electrical and Computer Engineering, University of Virginia, Charlottesville, VA 22904 USA
[4] Department of Material Science and Engineering, University of Virginia, Charlottesville, VA 22904 USA
* email: sjp9x@virginia.edu

(Dated: 25 October 2023)

The ability to manipulate spins in magnetic materials is essential in designing spintronics devices. One method for magnetic switching is through strain. In VO$_2$ on TiO$_2$ thin films, while VO$_2$ remains rutile across the metal-insulator transition, the in-plane lattice area expands going from low temperature insulating phase to high temperature conducting phase. In a VO$_2$/TbFeCo bilayer, the expansion of the VO$_2$ lattice area exerts tension on the amorphous TbFeCo layer. Through the strain effect, magnetic properties, including the magnetic anisotropy and magnetization, of TbFeCo can be changed. In this work, the changes in magnetic properties of TbFeCo on VO$_2$/TiO$_2$(011) are demonstrated using anomalous Hall effect measurements. Across the metal-insulator transition, TbFeCo loses perpendicular magnetic anisotropy, and the magnetization in TbFeCo turns from out-of-plane to in-plane. Using atomistic simulations, we confirm these tunable magnetic properties originating from the metal-insulator transition of VO$_2$. This study provides the groundwork for controlling magnetic properties through a phase transition

## I. INTRODUCTION

With the rapid developments of automation, the need for fast processing and compact data storage has promptly increased. Spintronic devices have the potential to serve as the building blocks of speedy data processors and high-density memory[1–6]. In spintronics, magnetic moments are the key components for reading and writing data. Being able to control magnetic moments is crucial in designing spintronic devices[3–6]. Several methods, such as current and laser pulses, can switch magnetic moments in multilayer thin films[7–9]. Investigating other methods to manipulate spins is critical for future developments in spintronics.

Among many mechanisms to control magnetism, straintronics, which employs strain-mediated effects for switching, presents an intriguing opportunity. It can serve as a foundation for energy-efficient devices[10–12]. One possibility of is using strain arises from the metal-insulator transition (MIT). For example, MIT in Vanadium dioxide (VO$_2$) has drawn interest from both fundamental theories and technological applications[13,14]. Recent studies have shown possible applications in ultrafast optics and electronic devices for sensing and switching[15–19]. In bulk VO$_2$, MIT occurs at 340K[20] and it is accompanied by abrupt changes in structural and electronic properties. Across MIT, bulk VO$_2$ undergoes a structural transition from a low-temperature monoclinic to a high-temperature rutile phase. In VO$_2$ thin films under uniaxial strain, recent reports reveal a complex mix of structural phases near MIT[21–27]. When VO$_2$ films are epitaxially grown on TiO$_2$ substrates, due to epitaxial bi-axial strains, the transitions are isostructural. In addition, MIT occurs at different temperatures, for VO$_2$ films grown on different orientations of the TiO$_2$ substrates. In VO$_2$/TiO$_2$, although VO$_2$ films remain rutile, the lattice parameters change along in-plane and out-of-plane directions[27]. Furthermore, in a similar V$_2$O$_3$ system, this coexistence of nanoscale phases near MIT leads to changes in magnetic properties in V$_2$O$_3$/Ni bilayers[28,29]. Moreover, magnetism in paramagnetic centers is found to be affected by MIT in VO$_2$ due to magnetoelastic anisotropy[30]. In these samples, the changes in lattice parameters of VO$_2$ serve as the most important mechanism for tuning magnetic properties. Because of their high magnetostrictions, ferrimagnetic rare-earth (RE) transitional-metal (TM) alloys such as TbFeCo are promising materials to study the effect on magnetism from MIT.

Amorphous ferrimagnetic RE-TM thin films have been widely studied for their applications in high-density low-current spintronics devices[31], sub-ps ultrafast magnetic switching[8,9,32–34], and a host for magnetic skyrmions with tunable Dzyaloshinskii-Moriya Interaction[35–39]. These ferrimagnetic films exhibit strong perpendicular magnetic anisotropy (PMA) and can be synthesized at room-temperature requiring no epitaxial growth[40,41]. Magnetic properties, such as magnetization and coercivity, are greatly influenced by the compensation temperature, which can be tuned by varying composition and thickness[42,43]. These properties make TbFeCo a good material to reveal the effect on magnetism from MIT.

In this work, the impact on magnetic properties from MIT is investigated in VO$_2$/TbFeCo bilayer. Amorphous TbFeCo films are grown on epitaxial VO$_2$ samples and Si/SiO$_2$ substrate. Comparison of magnetic properties reveals changes in magnetic anisotropy and magnetization in TbFeCo near MIT of VO$_2$. Furthermore, atomistic simulations are employed to incorporate the strain effect induced by VO$_2$ on TbFeCo near MIT. These results can serve as a foundation for devel-oping techniques to control magnetic properties through MIT for device applications. More importantly, since properties of VO$_2$[15–19] and RE-TM[8,9,32–34] can be controlled through an ultrafast laser, these results open up the possibility of high-speed data processing using RE-TM on VO$_2$.

## II. MATERIALS AND METHODS

~100 nm VO$_2$ thin films were grown on (011), and (100) TiO$_2$ substrates by reactive biased target ion beam deposition (RBTIBD). Details of growth conditions can be found in a previous publication[44]. 15 nm thick amorphous Tb$_{26}$Fe$_{64}$Co$_{10}$ thin films were deposited on VO$_2$/TiO$_2$ films and thermally oxidized Si substrates by RF magnetron sputtering at room temperature under base pressure of 5 x 10$^{-7}$ torr from co-sputtering of Tb and TbFeCo targets. The TbFeCo layers were deposited on the VO$_2$/TiO$_2$ films and SiO$_2$/Si substrates at the same time to eliminate changes in TbFeCo properties due to growth conditions. A 5 nm Ta capping layer was deposited on the samples to prevent oxidation. These samples were made in Hall bar devices for magneto-transport measurement, and Hall measurements were obtained for TbFeCo/VO$_2$/TiO$_2$(100), (011), and TbFeCo/SiO$_2$/Si samples.

Structural characterization of the samples was performed by X-ray diffraction (XRD) using a SmartLab system (Rigaku Inc.) in the 2$\theta$ range between 20 degrees and 80 degrees. Thin thickness measurements were performed by X-ray reflectivity (XRR) technique in the SmartLab. The film surface morphology was characterized via atomic force microscopy (AFM) by Cypher (Asylum Research Inc.). The magnetic properties at various temperatures were performed by vibrating sample magnetometer (VSM) option in a Versa Lab system (Quantum Design Inc.). The magneto-transport properties at various temperatures were performed by the electric transport option in the Versa Lab system. Temperatures were varied from 250 K to 350 K and applied magnetic fields were varied from -2 T to 2 T for these measurements.

| Parameter | Value |
|---|---|
| Fe Magnetic moment ($\mu_{Fe}$) | 2.22 $\mu_B$ |
| Tb Magnetic moment ($\mu_{Tb}$) | 9.34 $\mu_B$ |
| Fe-Fe Exchange Interaction ($J_{Fe-Fe}$) | 2.83 x 10$^{-21}$ J |
| Tb-Tb Exchange Interaction ($J_{Tb-Tb}$) | 0.99 x 10$^{-21}$ J |
| Fe-Tb Exchange Interaction ($J_{Fe-Tb}$) | -1.09 x 10$^{-21}$ J |
| Anisotropy ($K_u$) | 1 x 10$^5$ J/m$^3$ |
| Damping ($\alpha$) | 0.05 |

TABLE I. Values of parameters used in the atomistic simulations of TbFeCo.

An atomistic simulation was employed to study the change in magnetic hysteresis due to strain. A handmade atomistic code was used for the atomistic simulations. Since Fe and Co atoms belong to the same TM sublattice in the RE-TM ferrimagnet, Co atoms are treated as Fe atoms. Tb and Fe atoms are distributed in a 1.6 nm x 1.6 nm x 1.6 nm RE$_{25}$TM$_{75}$ amorphous structure. We placed replicas of this box next to each other in a 3 x 3 x 9 configuration to expand the simulation's size to 4.8 nm x 4.8 nm x 14.4 nm, and 20250 atoms in total. The parameters used in the simulation are listed in Table I. The anisotropy axis for each atom is distributed randomly within a 30-degree cone, with the axis of cone pointing along the out-of-plane direction. The exchange interactions are benchmarked based on Oslter et al.[45] and our experiments to maintain the same Curie temperature and compensation temperature for a given composition. Using stochastic Landau–Lifshitz–Gilbert (LLG) equation[46], hysteresis loops were simulated and compared to experiments. The strain anisotropy (K$_{strain}$) is given by

$$K_{strain} = -\frac{3}{2}\lambda E_y \varepsilon \quad (1)$$

where $\lambda$ = 100 ppm is the magnetostriction of amorphous TbFeCo, $E_y$ = 100 GPa is the Young's Modulus of TbFeCo and $\varepsilon$ is the strain exerted on TbFeCo by MIT of VO$_2$. In the case of TbFeCo thin films, a positive K$_{strain}$ leads to perpendicular magnetic anisotropy, while a negative K$_{strain}$ leads to in-plane magnetic anisotropy. The percentage of atoms that experience strain $\varepsilon$ varies with the phase distribution in VO$_2$ as VO$_2$ undergoes MIT, based on the fraction of metallic phase obtained from experiments. From Laverock et al.[26], VO$_2$'s transition is not abrupt across MIT. Near the MIT, there is a mixture of a low-temperature insulating phase and a high-temperature metallic phase present in the sample. To model this behavior, we approximated the fraction of atoms experiencing strain from VO$_2$'s MIT, based on the fraction of metallic phase obtained from the experiment by Laverock et al.[26] at various temperatures. For example, in TbFeCo on VO$_2$/TiO$_2$(011), no atoms experience strain at 250 K, 25% of atoms experience strain at 300 K and 75% of atoms experience strain at 320 K.

## III. RESULTS AND DISCUSSIONS

VO$_2$ films were grown on TiO$_2$ substrates with three different orientations. Fig. 1(a) presents XRD patterns of VO$_2$/TiO$_2$(011) (green), (100) (blue) films measured at room temperature. The 2$\theta$ peaks are indexed using rutile VO$_2$ (R-VO$_2$) and TiO$_2$. Different orientations of R-VO$_2$ are found in samples grown on different orientations of TiO$_2$ substrates. R-VO$_2$ (101), R-VO$_2$ (002), and R-VO$_2$ (200) peaks are observed in VO$_2$/TiO$_2$ (101), and VO$_2$/TiO$_2$ (100), respectively. These correspond to the epitaxial growth of VO$_2$ films in each TiO$_2$ orientation. The rutile phase in VO$_2$ at room temperature is consistent with the findings in a previous publication by Kittiwantanakul et al.[27]. In VO$_2$ thin films epitaxially grown on TiO$_2$ substrates, due to epitaxial bi-axial strains, VO$_2$ remains rutile in both the low-temperature insulating phase and the high-temperature conducting phase. Although VO$_2$ remains rutile, temperature-dependent XRD shows a change in relative lattice spacing across MIT. Above the MIT, the relative lattice space in VO$_2$ on TiO$_2$(100) becomes comparable to that of bulk VO$_2$[47]. Thus, in VO$_2$/TiO$_2$(011), the in-plane lattice area, defined as A = a x c expands from 12.66 Å$^2$, where a and
2

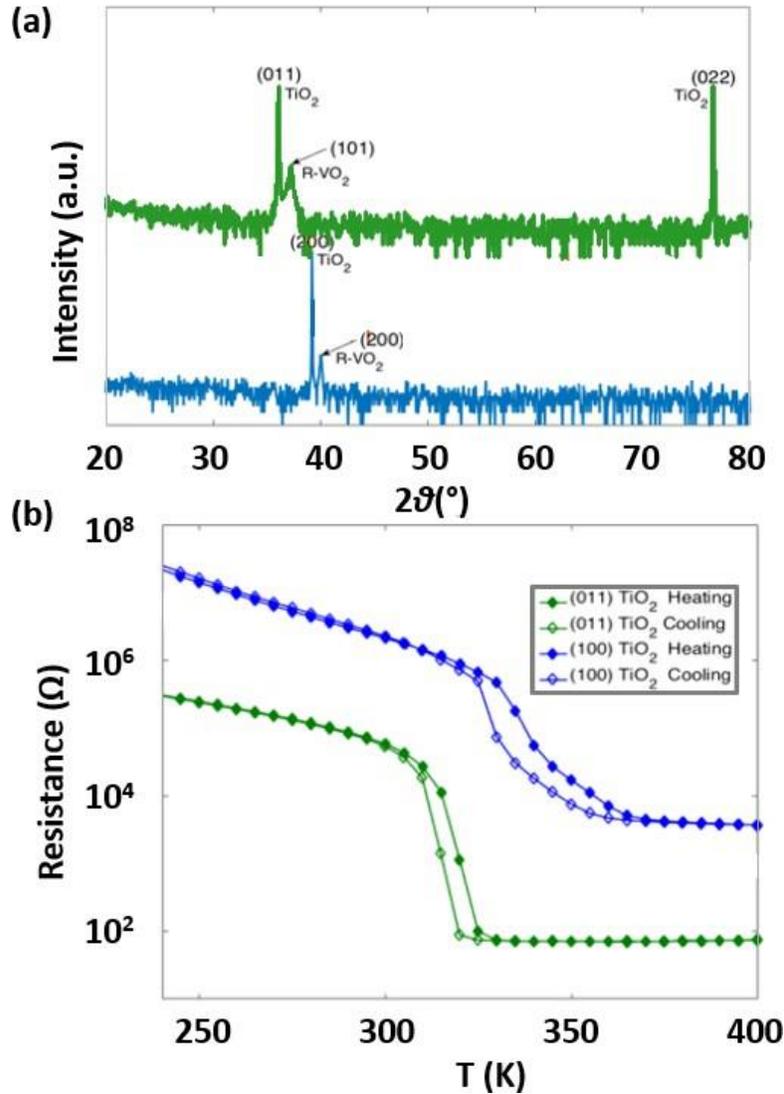

FIG. 1. (a) Room temperature X-ray diffraction (XRD) pattern of $VO_2/TiO_2(011)$ (green), and $VO_2/TiO_2(100)$ (blue) films. The $2\theta$ peaks are indexed with rutile $VO_2$ (R-$VO_2$) and $TiO_2$. (b) Resistance obtained from 240K to 400K in $VO_2/TiO_2(011)$ (green), and $VO_2/TiO_2(100)$ (blue). MIT of different orientations is observed at different temperatures between 310K and 350K.

c are lattice contants equal to 4.41Å and 2.87Å respectively, to 12.99 Å$^2$, where a and c are lattice contants equal to 4.56Å and 2.85Å respectively, going from the low-temperature insulating phase to the high-temperature conducting phase. On the other hand, in $VO_2/TiO_2(100)$, the in-plane lattice area compresses from 13.03 Å$^2$, where a and c are lattice contants equal to 4.51Å and 2.89Å respectively, to 12.99 Å$^2$, where a and c are lattice contants equal to 4.56Å and 2.85Å respectively, across MIT.

To characterize the MIT of $VO_2/TiO_2$, resistance measurements from 240 K to 400 K are shown in Fig. 1 (b). Across the MIT, $VO_2/TiO_2$ films show several orders of magnitude decrease in resistance, confirming the transition to a metallic state from an insulating state. Different orientations of $VO_2/TiO_2$ have different MIT temperatures between 310 K and 350 K. The MIT temperature is found in $VO_2/TiO_2$ (011) at ~320 K, followed by $VO_2/TiO_2(100)$ at ~350 K. Hysteresis-like behavior is present near MIT in all three orientations, where sharp changes in resistance occur at different temperatures under heating and cooling. The shift in MIT is due to different epitaxial bi-axial strains in $VO_2/TiO_2$ for different orientations[27].

To study the strain effect on magnetic properties from $VO_2$ across MIT, we deposited 15 nm thick TbFeCo with 5 nm thick Ta capping on top of various $VO_2/TiO_2$ films at the same time. Fig. 2 (a) shows a schematic diagram of the heterostructure investigated in this work. We studied the surface morphology and roughness in these films by AFM. Fig. 2 (b)-(e) show the AFM images of TbFeCo/$VO_2/TiO_2$ before and after the depositions of TbFeCo and Ta capping layer. Before the deposition of TbFeCo, the RMS roughnesses of the samples are 1.19 nm in $VO_2/TiO_2(011)$, and

0.66 nm in VO$_2$/TiO$_2$(100). After the deposition of TbFeCo and Ta capping layer, the RMS roughnesses of the samples are 1.29 nm in TbFeCo/VO$_2$/TiO$_2$(011), and 0.81 nm in TbFeCo/VO$_2$/TiO$_2$(100). This means that the changes in roughnesses after the deposition of TbFeCo are rather small. Furthermore, the AFM images show little changes to the samples' surfaces. These indicate the TbFeCo layers with Ta capping deposited on VO$_2$/TiO$_2$ maintained the same roughnesses and uniformity for each sample.

To investigate if there is any magnetic switching of TbFeCo due to MIT in VO$_2$, we fabricated each sample into Hall bar configurations and performed the anomalous Hall effect measurements on the patterned films. Anomalous Hall effect is considered here instead of direct hysteresis loops. This is because the TbFeCo films here have a low magnetization of about $1 \times 10^5$ A/m, resulting in a small magnetic moment signal in M-H loops measurements. Thus, anomalous Hall effect is considered here for clearer results from measurements. Fig. 3 (a)-(c) show normalized Hall resistance as a function of out-of-plane applied magnetic field of (a) TbFeCo/SiO$_2$/Si, (b) TbFeCo/VO$_2$/TiO$_2$(011), and (c) TbFeCo/VO$_2$/TiO$_2$(100). For higher temperatures, above 330 K, increases in noise are observed in both Fig. 3 (a) and (c). We suspect this is due to the temperature effect in the patterned films. In Fig. 3 (a), the Hall resistance of TbFeCo/SiO$_2$/Si shows very minor changes from 300 K to 350 K, which is expected. Since SiO$_2$/Si substrate has no transitions within this temperature range, the only source of strain acting on TbFeCo arises from the difference in thermal expansion between TbFeCo and SiO$_2$/Si substrate. The thermal expansion coefficient of SiO$_2$/Si substrate is 0.24 ppm/K. In comparison, the thermal expansion coefficient of amorphous TbFeCo near 300 K is about 10 ppm/K, estimated from amorphous TbFe alloy[48]. From 300 K to 350 K, $\varepsilon$ due to thermal expansion is $\sim$ 500 ppm, which is 5 x 10$^{-4}$. Using Eq. 1, this gives K$_{strain}$ of about -7.5 x 10$^3$ J/m$^3$, much smaller than K$_u$ of 1 x 10$^5$ J/m$^3$ in TbFeCo. As shown in Fig. 3 (a), $\varepsilon$ of 5 x 10$^{-4}$ is too small to have any effects on magnetic anisotropy in TbFeCo, and the magnetic moments of TbFeCo remain pointing in the out-of-plane directions at zero fields. These minor changes in hysteresis loops are likely due to an increase in temperature. The lack of significant changes in TbFeCo's out-of-plane loops shows that the magnetic anisotropy of TbFeCo is near constant around room temperature.

Next, we focus on the behavior of TbFeCo on VO$_2$/TiO$_2$ near room temperatures. From Fig. 3 (b), normalized Hall resistance as a function of out-of-plane applied magnetic field of TbFeCo/VO$_2$/TiO$_2$(011) shows a clear loss of PMA going from 250 K to 320 K. The magnetic hysteresis loops become less squared and the magnetic moments of TbFeCo switch from out-of-plane to in-plane. From Fig. 1 (b), the MIT of VO$_2$/TiO$_2$(011) (green line) occurs near 320 K. This means that the loss of PMA in TbFeCo corresponds to the MIT of VO$_2$ near 320 K. As the temperature goes up from 250 K to 320 K, VO$_2$'s in-plane lattice area expands across the MIT of VO$_2$/TiO$_2$(011). From Kittiwantanakul et al.[27], the in-plane lattice area expands from 12.66 Å$^2$ in the low-temperature phase to 12.99 Å$^2$ in the high-temperature phase. This corresponds to $\varepsilon$ of 2.6 x 10$^{-2}$ and K$_{strain}$ of -3.9 x 10$^5$ J/m$^3$ using Eq. 1, greater than K$_u$ of 1 x 10$^5$ J/m$^3$ in TbFeCo. Besides strain from VO$_2$'s in-plane lattice expansion, another source of strain arises from the difference in the thermal expansion between TbFeCo and VO$_2$. As discussed earlier, the thermal expansion coefficient of amorphous TbFeCo near 300 K is about 10 ppm/K. On the other hand, the thermal expansion coefficient of VO$_2$ near 300 K is about 21.1 ppm/K[49]. Thus, $\varepsilon$ due to thermal expansion going from 250 K to 320 K is $\sim$ 800 ppm, which is 8 x 10$^{-4}$. This is over an order of magnitude smaller than the $\varepsilon$ of 2.6 x 10$^{-2}$, arises from VO$_2$'s MIT. Moreover, from Laverock et al.[26], VO$_2$ films are not homogeneous. The MIT of VO$_2$ films involves a mixture of a low-temperature insulating phase and a high-temperature conducting phase across a temperature range. The means that TbFeCo on VO$_2$ is experiencing a gradual change in strain across MIT. This is supported by the progressive loss of PMA in TbFeCo going from 250 K to 320 K, as seen in Fig. 3 (b). This shows that the switching of TbFeCo from out-of-plane to in-plane is likely due to the tensile strain that arises from VO$_2$'s in-plane lattice expansion across MIT.

Fig. 3 (c) shows the normalized Hall resistance as a function out-of-plane applied magnetic field of TbFeCo/VO$_2$/TiO$_2$(100). The Hall effect of TbFeCo/VO$_2$/TiO$_2$(100) reveals the absence of PMA in TbFeCo throughout the measured temperature. This is probably due to the presence of tensile strain acting on TbFeCo by VO$_2$/TiO$_2$(100). The in-plane lattice area of the low-temperature insulating phase in VO$_2$/TiO$_2$(100) is 13.03 Å$^2$[27], which is larger compared to the in-plane lattice area in VO$_2$/TiO$_2$(011) (12.66 Å$^2$). This means that the underlayer of VO$_2$/TiO$_2$(100) is most likely applying a tensile interfacial strain on the TbFeCo atoms in these multilayer thin films. Since amorphous TbFeCo has positive magnetostriction, a tensile strain will lead to an additional in-plane anisotropy contribution. In contrast, in TbFeCo/VO$_2$/TiO$_2$(011), the smaller in-plane lattice area at the low-temperature insulating phase in VO$_2$/TiO$_2$(011) is creating a compressive interfacial strain on TbFeCo, resulting in PMA in TbFeCo. Furthermore, when the in-plane lattice area of VO$_2$/TiO$_2$(011) expands to 12.99 Å$^2$ across the MIT, TbFeCo on VO$_2$/TiO$_2$(011) lost PMA. This shows that the in-plane lattice area of 12.99 Å$^2$ or larger is supplying a tensile interfacial strain on TbFeCo.

From Fig. 3 (c), as the temperature changes from 300 K to 350 K, there are no changes in the magnetic anisotropy of TbFeCo, magnetic moments of TbFeCo remain in-plane at zero external fields throughout the measured temperatures. From Fig. 1 (b), the MIT of VO$_2$/TiO$_2$(100) (blue line) occurs near 350 K. This means that across the MIT of VO$_2$, magnetic properties of TbFeCo remain unaffected. This can be explained by the change in the in-plane lattice area of VO$_2$/TiO$_2$(100) across MIT. In VO$_2$/TiO$_2$(100), the in-plane lattice area shrank from 13.03 Å$^2$ in the low-temperature phase to 12.99 Å$^2$ in the high-temperature phase[27]. This corresponds to an $\varepsilon$ of -3 x 10$^{-3}$. Note that the negative sign here corresponds to the compressive strain exerted on TbFeCo, compared to tensile strain in the other samples. The strain in VO$_2$/TiO$_2$(100) is almost 10 times smaller than the strain in VO$_2$/TiO$_2$(011), which is 2.6 %. Therefore, it makes sense that the change in magnetic anisotropy of TbFeCo is only observed in TbFeCo/VO$_2$/TiO$_2$(011), but not in TbFeCo/VO$_2$/TiO$_2$(100).

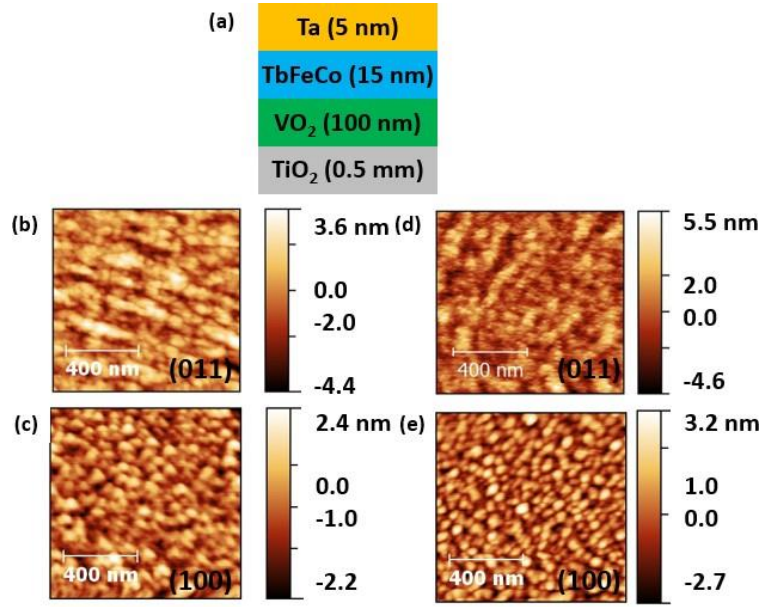

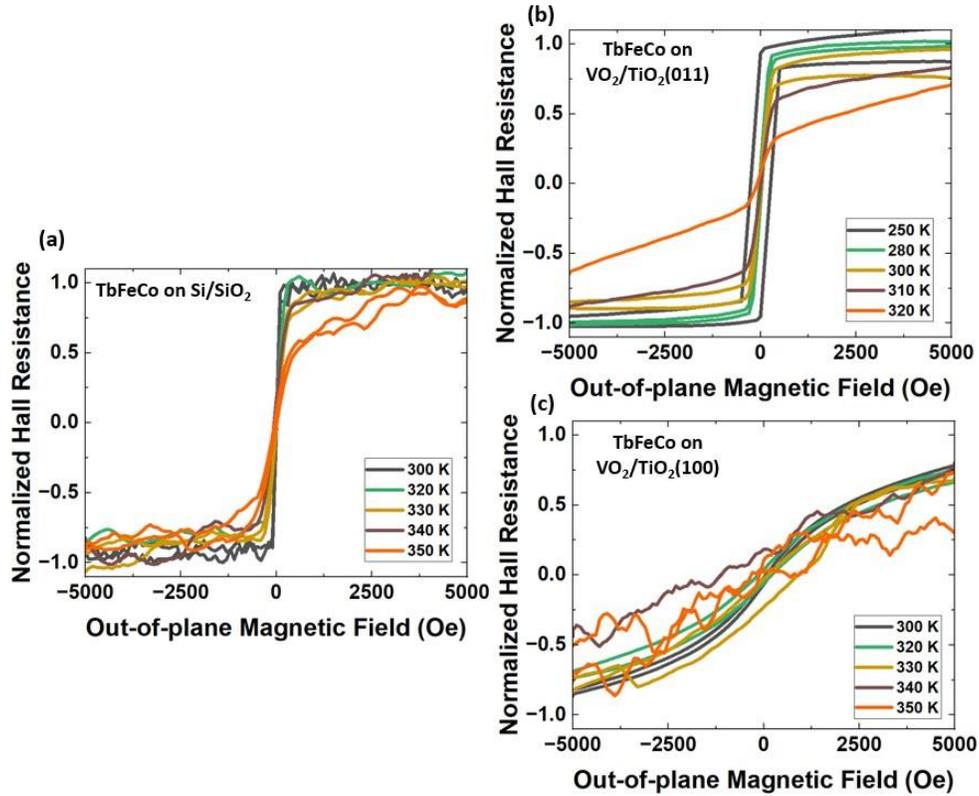

FIG. 2. (a) An illustration of the TbFeCo/VO$_2$ heterostructure (not to scale). (b-e) Atomic force microscopy (AFM) images of TbFeCo/VO$_2$/TiO$_2$ (b-c) before and (d-e) after the deposition of TbFeCo layer with Ta capping layer, (b) VO$_2$/TiO$_2$(011); (c) VO$_2$/TiO$_2$(100); (d) TbFeCo/VO$_2$/TiO$_2$(011); (e) TbFeCo/VO$_2$/TiO$_2$(100).

FIG. 3. Anomalous Hall effect of TbFeCo measured at various temperatures under an out-of-plane external field. (a) TbFeCo/SiO$_2$/Si; (b) TbFeCo/VO$_2$/TiO$_2$(011); (c) TbFeCo/VO$_2$/TiO$_2$(100).

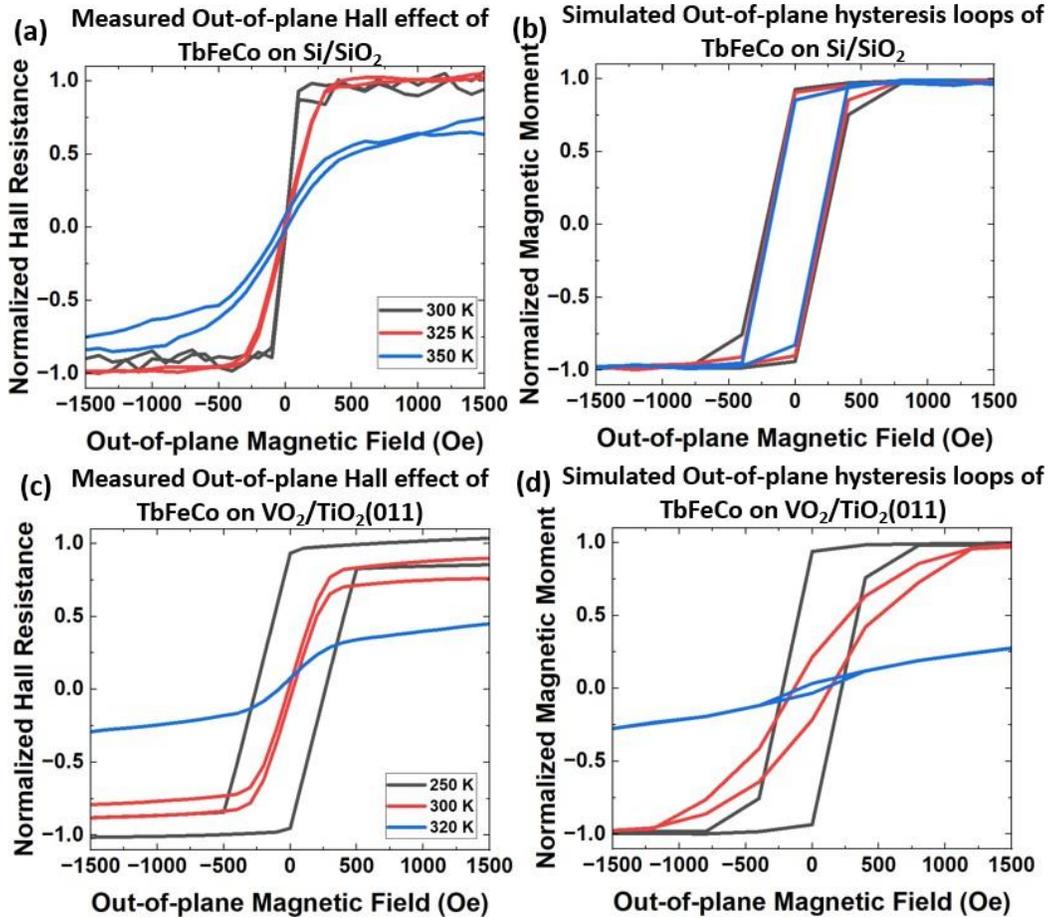

FIG. 4. Comparison of (a) measured out-of-plane anomalous Hall effect (extracted from Fig. 3 (a)) and (b) simulated out-of-plane hys- teresis loops at various temperatures in TbFeCo/SiO$_2$/Si by atomistc simulations. Comparison of (c) measured out-of-plane anomalous Hall effect (extracted from Fig. 3 (b)) and (d) simulated out-of-plane hysteresis loops at various temperatures with strain anisotropy in TbFeCo/VO$_2$/TiO$_2$(011) by atomistc simulations.

To verify that the strain from VO$_2$'s MIT is the source of magnetic switching in TbFeCo, an atomistic model is employed. In this model, strain anisotropy is given by Eq. 1. Fig. 4 (a) and (b) show the comparison of measured out-of-plane anomalous Hall effect and simulated hysteresis loops from 300 K to 350 K in TbFeCo on SiO$_2$/Si substrate, respectively. In this sample, no strain anisotropy is included in the simulations because SiO$_2$/Si substrate does not undergo transition across these temperatures. Results indicate the minor changes in measured anomalous Hall effect from 300 K to 350 K are due to an increase in temperature. A discrepancy in the coercivity between measurements and simulations is observed in Fig. 4 (a) and (b). We suspect the discrepancy originates from the complex cone-shaped anisotropy in amorphous rare-earth transition-metal films[40]. Next, we investigated hysteresis loops of TbFeCo on VO$_2$/TiO$_2$(011) using atomistic simulations. Fig. 4 (c) and (d) show the comparison of measured out-of-plane anomalous Hall effect and simulated out-of-plane hysteresis loops from 250 K to 320 K in TbFeCo/VO$_2$/TiO$_2$(011), respectively. With the incorporated model of the strain anisotropy, the measured and simulated hysteresis loops are in good agreement. Both show the gradual loss of PMA in TbFeCo from 250 K to 320 K and the magnetic moments turn from out-of-plane to in-plane at zero fields going from 250 K to 320 K. This confirms that strain from VO$_2$'s MIT is the source of magnetic switching in TbFeCo.

## IV. CONCLUSIONS

In summary, 15 nm thick amorphous TbFeCo films were deposited on VO$_2$/TiO$_2$ to study the strain effect of metal-insulator transition (MIT) on magnetic properties. Using TbFeCo on thermally oxidized Si substrate as a reference sample, changes in magnetic anisotropy were observed in TbFeCo/VO$_2$/TiO$_2$(011) film. Near the MIT of VO$_2$/TiO$_2$(011), a decrease in magnetic anisotropy was found in TbFeCo and the magnetization of TbFeCo switched from out-of-plane to in-plane at zero external fields. This decrease in magnetic anisotropy originated from the tensile strain arising from the transition of $VO_2/TiO_2(011)$, where the in-plane lattice area of $VO_2$ expands. Furthermore, atomistic simulations of TbFeCo with strain anisotropy from $VO_2$ were in agreement with measurements, confirming that the in-plane lattice expansion in $VO_2/TiO_2(011)$ across MIT is sufficient to switch magnetic moments in TbFeCo. These results offer a platform for using the phase transition to achieve magnetic switching in spintronics devices for desirable applications.

C.T.M, S.K, A.S, and Y.W.: sample fabrication, and measurements. C.T.M and M.G.M: modeling, writing-review, and editing. A.W.G and S.J.P.: supervision and writing-review. All authors have read and agreed to the published version of the manuscript.

This research received no external funding.

The data that support the findings of this study are available from the corresponding author upon reasonable request.

The authors declare no conflict of interest.

**ACKNOWLEDGMENTS**

We thank Dr. Jiwei Lu for his thoughtful and stimulating discussions. S.K. acknowledges the support from the NSRF via the Program Management Unit for Human Resources & Institutional Development, Research and Innovation [grant number B05F650024].